%% file: main.tex
\documentclass[11pt]{article}
\usepackage{tcolorbox}
\usepackage[utf8]{inputenc}
\usepackage{algorithm}
\usepackage{algorithmic}
\usepackage[letterpaper,hmargin=1.0in,vmargin=1.0in]{geometry}
\usepackage{natbib}
\usepackage{subfigure}
\usepackage{mwe}

\def\fairletDec{FairletDecomposition}

\def\unbalancedPoints{UnbalancedPoints}
\def\extraPoints{ExtraPoint}
\def\freeFairlet{NonSaturFairlet}

\def\dom{\mathrm{dom}}

\def\minHeavy{MinHeavyPoints}

\def\clusterFairlet{ClusterFairlet}

\def\heavy{\mathrm{H}}

\def\med{\mathrm{median}}

\def\smed{\mathrm{med}}

\def\lca{\textsf{lca}}
\def\bal{\mathbf{balance}}

\def\sC{\mathcal{C}}

\usepackage{xparse}
\NewDocumentCommand\Call{m+g}{%
  \IfNoValueTF{#2}
    {\textsc{#1}}
    {\textsc{#1}(#2)}%
}

\include{prefix}
\title{Scalable Fair Clustering}

\author{
Arturs Backurs\thanks{TTIC. {\tt backurs@ttic.edu}} 
\and
Piotr Indyk\thanks{CSAIL, MIT. {\tt\{indyk, vakilian, talw\}@mit.edu}}
\and
Krzysztof Onak\thanks{MIT-IBM Watson AI Lab, IBM Research. {\tt konak@us.ibm.com}}
\and
Baruch Schieber\thanks{Department of Computer Science, New Jersey Institute of Technology. {\tt baruch.m.schieber@njit.edu}}
\and
Ali Vakilian\footnotemark[2]
\and
Tal Wagner\footnotemark[2]
}	

\begin{document}
\setlength{\abovedisplayskip}{3pt}
\setlength{\belowdisplayskip}{3pt}

\thispagestyle{empty}%
\setcounter{page}{0}

\pagenumbering{gobble}
\date{}
\maketitle{}
\setcounter{page}{1}%
\pagenumbering{arabic}%

\begin{abstract}
We study the fair variant of the classic $k$-median problem introduced by~\cite{chierichetti2017fair}. In the standard $k$-median problem, given an input pointset $P$, the goal is to find $k$ centers $C$ and assign each input point to one of the centers in $C$ such that the average distance of points to their cluster center is minimized.
 In the fair variant of $k$-median, the points are colored, and the goal is to minimize the same average distance objective while ensuring that all clusters have an ``approximately equal'' number of points of each color. 

\cite{chierichetti2017fair} proposed a two-phase algorithm for fair $k$-clustering. In the first step, the pointset is partitioned into subsets called {\em fairlets} that satisfy the fairness requirement and approximately preserve the $k$-median objective. In the second step,  fairlets are merged into $k$ clusters by one of the existing $k$-median algorithms. The running time of this algorithm is dominated by the first step, which takes super-quadratic time. 

In this paper, we present a practical approximate fairlet decomposition algorithm that runs in {\em nearly linear} time. Our algorithm additionally allows for finer control over the balance of resulting clusters than the original work. We complement our theoretical bounds with empirical evaluation.
\end{abstract}

\input{intro}

\input{prelim}

\input{alg_description}
\input{fair_clustering}

\input{experiment}

\section*{Acknowledgment}
The authors would like to thank Ravi Kumar for many helpful discussions. This project was supported by funds from the MIT-IBM Watson AI Lab, NSF, and Simons Foundation. 

\appendix
\input{missing_proofs}
\end{document}

%% file: prefix.tex
\usepackage{bm}
\usepackage{amsfonts,amsmath,amssymb}
\usepackage{graphicx}
\usepackage{xspace}
\usepackage{boxedminipage}
\usepackage{framed}
\usepackage{url}
\usepackage{float}
\usepackage{enumerate,paralist}
\usepackage{accents}
\usepackage{setspace}
\usepackage{hyperref}
\usepackage{mathrsfs}
\usepackage[amsmath,thmmarks]{ntheorem}%
\usepackage{graphicx}
\usepackage{colortbl}
\usepackage{hhline}
\usepackage{tikz}
\usepackage{multirow}
\usetikzlibrary{decorations.pathreplacing,angles,quotes}
\usepackage[normalem]{ulem}
\usepackage{afterpage}
\usepackage{wrapfig}
\theoremseparator{.}%

\theoremstyle{plain}%
\newtheorem{theorem}{Theorem}[section]

\newtheorem{lemma}[theorem]{Lemma}
\newtheorem{corollary}[theorem]{Corollary}
\newtheorem{claim}[theorem]{Claim} 
\newtheorem{question}[theorem]{Question}
\newtheorem{definition}[theorem]{Definition}

\theoremheaderfont{\em}%
\theorembodyfont{\upshape}%
\theoremstyle{nonumberplain}%
\theoremseparator{}%
\theoremsymbol{\myqedsymbol}%
\newtheorem{proof}{Proof:}%

\newcommand{\myqedsymbol}{$\square$}

\newtheorem{proofof}{Proof of\!}%

\newcommand{\eps}{\varepsilon}%
\def\bar{\overline}
\def\floor#1{\lfloor {#1} \rfloor}
\def\ceil#1{\lceil {#1} \rceil}
\def\script#1{\mathcal{#1}}
\def\opt{\textsc{OPT}}
\def\etal{\text{et al.}\xspace}

\def\card#1{|#1|}
\def\set#1{\{#1\}}

\newcommand{\sol}{\textsc{sol}}%

\newcommand{\argmax}{\mathrm{argmax}{}\xspace}

\newcommand{\poly}{\mathop{\mathrm{poly}}}%

\def\cost{\mathrm{cost}}

\newcommand{\tldO}{\widetilde{O}}%

%% file: intro.tex
\section{Introduction}

The success of machine learning led to its widespread adoption in many aspects of our daily lives. Automatic prediction and forecasting methods are now used to approve mortgage applications or estimate the likelihood of recidivism~\citep{chouldechova2017fair}. It is thus crucial to design machine learning algorithms that are {\em fair}, i.e., do not suffer from bias against or towards particular population groups. An extensive amount of research over the last few  years has focused on two key questions: how to formalize the notion of fairness in the context of common machine learning tasks, and how to design efficient algorithms that conform to those formalizations. See e.g., the survey by \cite{chouldechova2018frontiers} for an overview.

In this paper we focus on the second aspect. Specifically, we consider the problem of {\em fair clustering} and propose efficient algorithms for solving this problem. Fair clustering, introduced in~\citep{chierichetti2017fair}, generalizes the standard notion of clustering by imposing a constraint that all clusters must be {\em balanced} with respect to specific sensitive attributes, such as gender or religion. In the simplest formulation, each input point is augmented  with one of two colors (say, red and blue), and the goal is to cluster the data while ensuring that, in each cluster, the fraction of points with the less frequent color is bounded from below by some parameter strictly greater than $0$. Chierichetti \etal proposed polynomial time approximation algorithms for fair variants of classic clustering methods, such as $k$-center (minimize the {\em maximum} distance between points and their cluster centers) and $k$-median (minimize the {\em average} distance between points and their cluster centers). To this end, they introduced the notion of {\em fairlet decomposition}: a partitioning of the input pointset into small subsets, called {\em fairlets}, such that a good balanced clustering can be obtained by merging fairlets into clusters. Unfortunately, their algorithm for computing a fairlet decomposition has running time that is at least {\em quadratic} in the number of the input points. As a result, the algorithm is applicable only to relatively small data sets.

In this paper we address this drawback and propose an algorithm for computing fairlet decompositions with running time that  is {\em near-linear} in the data size. We focus on the $k$-median formulation, as $k$-center clustering is known to be sensitive to outliers. Our algorithms apply to the typical case where the set of input points lie in a $d$-dimensional space, and the distance is induced by the Euclidean norm.\footnote{E.g., all data sets used to evaluate the algorithms in~\citep{chierichetti2017fair} fall into this category.}

To state the result formally, we need to introduce some notation. Consider a collection of $n$ points $P \subseteq \mathbb{R}^d$, where each point $p \in P$ is colored either {\em red} or {\em blue}. For a subset of points $S\subseteq P$, the {\em balance} of $S$ is defined as $\bal(S) := \min\set{{\card{S_r} \over \card{S_b}}, {\card{S_b} \over \card{S_r}}}$ where $S_r$ and $S_b$ respectively denote the subset of red and blue points in $S$.
Assuming $b<r$, a clustering $\sC=\{C_1 \ldots C_k\}$  of $P$ is {\em ($r,b$)-fair} if for every cluster $C\in \sC$, $\bal(C) \geq {b\over r}$. 
In $k$-median clustering, the goal is to find $k$ centers and partition the pointset $P$ into $k$ clusters centered at the selected centers such that the sum of the distances from each point $p\in P$ point to its assigned center (i.e., the center of the cluster to which $p$ belongs) is minimized. In the $(r,b)$-fair $k$-median problem, all clusters are required to have balance at least ${b\over r}$.
Our main result is summarized in the following theorem. 
\begin{theorem}\label{thm:main-fair-k-median}
Let $T(n,d,k)$ be the running time of an $\alpha$-approximation algorithms for the $k$-median problem over $n$ points in $\mathbb{R}^d$. 
Then there exists an $O(d\cdot n \cdot \log n + T(n,d,k))$-time algorithm that given a point set $P\subseteq \mathbb{R}^d$ and balance parameters $(r,b)$, computes a $(r,b)$-fair $k$-median of $P$ whose cost is within a factor of $O_{r,b}(d \cdot \log n + \alpha)$  from the optimal  cost of $(r,b)$-fair $k$-median of $P$.
\end{theorem}

The running time can be reduced further by applying dimensionality reduction techniques, see, e.g., \citep{cohen2015dimensionality, makarychev2018performance} and the references therein.

We complement our theoretical analysis with empirical evaluation. Our experiments show that the quality of the clustering obtained by our algorithm is comparable to that of~\cite{chierichetti2017fair}. At the same time, the empirical runtime of our algorithm scales almost linearly in the number of points, making it applicable to massive data sets (see Figure~\ref{fig:runtime}).  

\paragraph{Related work.} Since the original paper of 
\cite{chierichetti2017fair}, there has been several followup works studying fair clustering. In  particular, \cite{rosner18privacy} and \cite{bercea2018cost} studied the fair variant of $k$-center clustering (as opposed to $k$-median in our case). Furthermore, the latter paper presented a ``bi-criteria'' approximation algorithm for $k$-median and $k$-means under a somewhat different notion of fairness. However, their solution relies on a  linear program that is a relaxation of an integer linear program with at least $n^2$ variables, one for every pair of points. Thus, their algorithm does not scale well with the input size.
Another algorithm proposed in \cite{fairclustering19}, requires solving a linear program with $nk$ variables. Due to the special structure of the LP it is plausible that it can be solved efficiently, but we are not aware of any empirical evaluation of this approach.


The work most relevant to our paper is a recent manuscript by~\cite{schmidt2018fair}, which proposed efficient streaming algorithms for fair $k$-means (which is similar to  $k$-median studied here). Specifically, they give a near-linear time streaming algorithm for computing a {\em core-set}: a small subset $S \subseteq P$ such that solving fair clustering over $S$ yields an approximate solution for the original point-set $P$. 
In order to compute the final clustering, however, they still need to apply a fair clustering algorithm to the core-set. Thus, our approach is complementary to the core-set approach, and the two can be combined to yield algorithms which are both fast and space-efficient\footnote{We note, however, that since core-sets typically require assigning weights to data points, such combination requires extending the clustering algorithm to weighted pointsets. In this paper we do not consider the weighted case.}.

We note that the above algorithms guarantee constant approximation factors, as opposed to the  logarithmic factor in our paper. As we show in the experimental section, this does not seem to affect the empirical quality of solutions produced by our algorithm. Still, designing a constant factor algorithm with a near-linear running time is an interesting open problem.

\paragraph{Possible settings of $(r,b)$.}
\cite{chierichetti2017fair} gave $(r,b)$-fairlet decomposition algorithms only for $b=1$. This does not allow for computing a full decomposition of the pointset into well-balanced fairlets if the numbers of red and blue points are close but not equal (for instance, if their ratio is 9:10). One way to address this could be to downsample the larger set in order to make them have the same cardinality and then compute a $(1,1)$-fairlet decomposition. The advantage of our approach is that we do not disregard any, even random, part of the input. This may potentially lead to much better solutions, partially by allowing that the clusters are not ideally balanced. The general settings of $r$ and $b$ are also considered by~\cite{bercea2018cost, fairclustering19}.

\paragraph{Our techniques.} Our main contribution is to design a nearly-linear time algorithm for $(r,b)$-fairlet decomposition for any integer values of $r,b$. 
Our algorithm has two steps. First, it embeds the input points into a  tree metric called {\em HST} (intuitively, this is done by computing a quadtree decomposition of the point set, and then using the distances in the quadtree).  In the second step it  solves the fairlet decomposition problem with respect to the new distance function induced by HST. The distortion of the  embedding into the HST accounts for the $\log n$ factor in the approximation guarantee. 

Once we have the HST representation of the pointset, the high-level goal is to construct ``local'' $(r,b)$-fairlets with respect to the tree. To this end, the algorithm scans the tree in a top-down order. In each node $v$ of the tree, it greedily partitions the points into fairlets so that the number of fairlets whose points belong to subtrees rooted at different children of $v$ is minimized. In particular, we prove that minimizing the number of such fairlets (which we refer to as the \emph{Minimum Heavy Point} problem) leads to an $O(1)$-approximate $(r,b)$-fairlet decomposition with respect to the distance over the tree.

%% file: prelim.tex
\section{Preliminaries}\label{sec:prelim}

\begin{definition}[Fairlet Decomposition]\label{def:fairlet-dec}
Suppose that $P\subseteq Y$ is a collection of points such that each is either colored {\em red} or {\em blue}. Moreover, suppose that $\bal(P) \geq {b\over r}$ for some integers $1\leq b\leq r$ such that $\mathrm{gcd}(r,b)=1$. A clustering $\script{X} = \set{D_1, \cdots, D_m}$ of $P$ is an \emph{$(r,b)$-fairlet decomposition} if \emph{(a)} each point $p\in P$ belongs to exactly one cluster $D_i \in \script{X}$, \emph{(b)} for each $D_i\in \script{Y}$, $\card{D_i} \leq b+r$, and \emph{(c)} for each $D_i \in \script{X}$, $\bal(D_i) \geq {b\over r}$. 
\end{definition}

\paragraph{Probabilistic metric embedding.}
A probabilistic metric $(X, \bar{d})$ is defined as a set of $\ell$ metrics $(X, d_1), \cdots, (X, d_\ell)$ along with a probability distribution of support size $\ell$ denoted by $\alpha_1, \cdots, \alpha_\ell$ such that $\bar{d}(p,q) = \sum_{i=1}^\ell \alpha_i \cdot d_i(p,q)$. 
For any finite metric $M = (Y,d)$ and probabilistic metric $(X, \bar{d})$, an {\em embedding} $f: Y \rightarrow X$ has distortion $c_f$, if:
\begin{itemize}
\item{for all $p,q \in Y$ and $i \leq k$, $d_i(f(p), f(q)) \geq d(p,q)$,}
\item{$\bar{d}(f(p), f(q)) \leq c_f \cdot d(p,q)$.}
\end{itemize}

\begin{definition}[$\gamma$-HST]\label{def:HST}
A tree $T$ rooted at vertex $r$ is a \emph{hierarchically well-separated tree ($\gamma$-HST)} if all edges of $T$ have non-negative weights and the following two conditions hold:
\begin{enumerate}
\item{The (weighted) distances from any node to all its children are the same.}
\item{For each node $v \in V\setminus\set{r}$, the distance of $v$ to its children is at most ${1/\gamma}$ times the distance of $v$ to its parent.}
\end{enumerate}
\end{definition}

We build on the following result due to~\cite{bartal1996probabilistic}, which gives a probabilistic embedding from $\mathbb{R}^d$ to a $\gamma$-HST. Our algorithm explicitly computes this embedding which we describe in more detail in the next section. For a proof of its properties refer to~\citep{indyk2001algorithmic} and the references therein. In this paper, we assume that the given pointset has $\poly(n)$ {\em aspect ratio} (i.e. the ratio between the maximum and minimum distance is $\poly(n)$).

\begin{theorem}[\cite{bartal1996probabilistic}]\label{thm:tree-metric-distortion}
Any finite metric space $M$ on points in $\mathbb{R}^d$ can be embedded into probabilistic metric over $\gamma$-HST metrics with $O(\gamma\cdot d \cdot \log_\gamma n)$ distortion in $O(d\cdot n \cdot \log_{\gamma} n)$ time.
\end{theorem}

%% file: alg_description.tex
\section{High-level Description of Our Algorithm}\label{sec:high-level}
Our algorithm for $(r,b)$-fair $k$-median problem in Euclidean space follows the high-level approach of 
\cite{chierichetti2017fair}: it first computes an approximately optimal $(r,b)$-fairlet decomposition for the input point set $P$ (see Algorithm~\ref{alg:recursive-fairlet-decomposition}). Then, in the second phase, it clusters the ($r,b$)-fairlets produced in the first phase into $k$ clusters (see Algorithm~\ref{alg:fairlet-to-clusters}). Our main contribution is designing a \emph{scalable} algorithm for the first phase of this approach, namely {\em ($r,b$)-fairlet decomposition}.

\begin{figure*}[!h]
\centering
\subfigure[The original space partitioned into hypercubes.]{\includegraphics[width=0.27\textwidth]{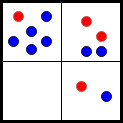}}
\quad
\subfigure[A 2-HST embedding of the input points.]{\includegraphics[width=0.27\textwidth]{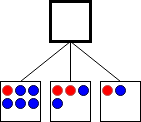}}
\quad 
\subfigure[Stage 1: we must connect 3 blue points from the left node through the root.]{\includegraphics[width=0.27\textwidth]{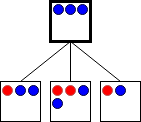}}
\\
\centering
\subfigure[Stage 2: we can connect 1 red point from the middle node through the root.]{\includegraphics[width=0.27\textwidth]{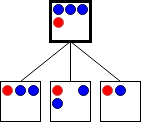}}
\quad
\subfigure[Stage 3: we add the unsaturated fairlet in the right node to the root and make it balanced.]{\includegraphics[width=0.27\textwidth]{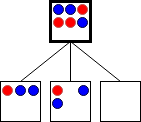}}
\quad
\subfigure[The final fairlet clustering.]{\includegraphics[width=0.27\textwidth]{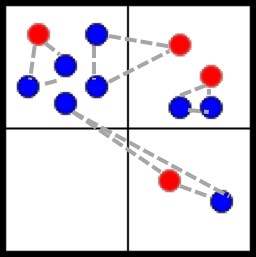}}
\caption{A run of our algorithm for (1,3)-fairlet decomposition on 8 blue points and 4 red points in $\mathbb{R}^2$. Steps (c)-(e) show the three stages of step 1 in \Call{\fairletDec}.}
\label{fig:alg-description}
\end{figure*}

\paragraph{Preprocessing phase: embedding to $\gamma$-HST.}
An important step in our algorithm is to embed the input pointset $P$ into a $\gamma$-HST (see Section~\ref{sec:prelim} for more details on HST metrics). To this end, we exploit the following standard construction of $\gamma$-HST using {\em randomly shifted grids}. 

Suppose that all points in $P$ lie in $\set{-\Delta, \cdots, \Delta}^d$. We generate a random tree $T$ (which is a $\gamma$-HST embedding of $P$) recursively. We translate the $d$-dimensional hypercube $H = [-2\Delta,2\Delta]^d$ via a uniformly random shift vector $\sigma\in \set{-\Delta,\cdots,\Delta}^d$. 
It is straightforward to verify that all points in $P$ are enclosed in $H + \sigma$. 
We then split each dimension of $H$ into $\gamma$ equal pieces to create a grid with $\gamma^d$ cells. 
Then we proceed recursively with each non-empty cell to create a hierarchy of nested $d$-dimensional grids with $O(\log_{\gamma} {\Delta \over \eps})$ levels (each cell in the final level of the recursion either contains exactly one point of $P$ or has side length $\eps$). 
Next, we construct a tree $T$ corresponding to the described hierarchy nested $d$-dimensional grids as follows. Consider a cell $C$ in the $i$-th level (level $0$ denote the initial hypercube $H$) of the hierarchy. 
Let $T^{1}_C, \cdots, T^{\ell}_C$ denote the trees constructed recursively for each non-empty cells of $C$. 
Denote the root of each tree $T^{j}_C$ by $u^{j}_C$. 
Then we connect $u_C$ (corresponding to cell $C$) to each of $u^{C}_j$ with an edge of length proportional to the diameter of $C$ (i.e., $(\sqrt{d} \cdot \Delta)/\gamma^i$). 

Note that the final tree generated by the above construction is a $\gamma$-HST: on each path from the root to a leaf, the length of consecutive edges decrease exponentially (by a factor of $\gamma$) and the distance from any node to all of its children are the same. Moreover, we assume that $\Delta/\eps = n^{O(1)}$.



\paragraph{Phase 1: computing ($r,b$)-fairlet decomposition.}  
This phase operates on the probabilistic embedding of the input into a $\gamma$-HST $T$ from the preprocessing phase, where $\gamma = \poly(r,b)$. The distortion of the embedding is $O(d\cdot \gamma \cdot \log_{\gamma} n)$. 
%
Additionally, we augment each node $v\in T$ with integers $N_r$ and $N_b$ denoting the number of {\em red} and {\em blue} points, respectively, in the subtree $T(v)$ rooted at $v$. 
%

\begin{algorithm}[!h]
\caption{\Call{\fairletDec}{$v, r, b$}: returns an ($r,b$)-fairlet decomposition of the points in $T(v)$}
\label{alg:recursive-fairlet-decomposition}
\begin{algorithmic}[1]
	
	\IF{$v$ is a leaf node of $T$}
		\RETURN an arbitrary ($r,b$)-fairlet decomposition of the points in $T(v)$ 
	\ENDIF

	\item[]
	\COMMENT{{\bf Step 1:} approximately minimize the total number of heavy points with respect to $v$}
	\STATE $\set{x^i_r, x^i_b}_{i} \leftarrow \Call{\minHeavy}{\set{N_r^i, N_b^i}_{i\in [\gamma^d]}, r, b}$ \COMMENT{for non-empty children $i\in[\gamma^d]$ of $v$}
	
	\item[]
	\COMMENT{{\bf Step 2:} find an ($r,b$)-fairlet decomposition of heavy points with respect to $v$}
	\STATE $P_v \leftarrow \emptyset$
	\FORALL{non-empty children $i\in[\gamma^d]$ of $v$}
		\STATE {\bf remove} an arbitrary set of $x^i_r$ red and $x^i_b$ blue points from $T(v_i)$ and {\bf add} them to $P_v$ 
	\ENDFOR
	\STATE {\bf output} an $(r,b)$-fairlet decomposition of $P_v$
	
	\item[]
	\COMMENT{{\bf Step 3:} proceed to the children of $v$}
	\FORALL{non-empty children $i\in[\gamma^d]$ of $v$}
		\STATE \Call{\fairletDec}{$v_i, r, b$} 
	\ENDFOR
\end{algorithmic}
\end{algorithm}

\paragraph{Step 1.} Compute an {\em approximately minimum} number of points that are required to be removed from the children of $v$ so that (1) the set of points contained by each child becomes $(r,b)$-balanced, and (2) the union of the set of removed points is also $(r,b)$-balanced.  
More formally, we solve Question~\ref{q:min-heavy-points} approximately (recall that for each child $v_i$, $N^i_r$ and $N^i_b$ respectively denotes the number of red and blue points in $T(v_i)$).
\begin{definition}[{\bf Heavy Point}]\label{def:heavy-points}
A point $p\in T(v)$ is {\em heavy} with respect to $v$ if it belongs to a fairlet $D$ such that $\lca(D) = v$.
For each fairlet $D\in \script{X}$, $\lca(D)$ denotes the {\em least common ancestor (lca)} of the points contained in $D$ in $T$.
\end{definition}
\begin{question}[Minimum Heavy Points Problem]\label{q:min-heavy-points}
Suppose that $v$ is a node in $T$. For each $i\in [\gamma^d]$ corresponding to non-empty children of $v$, let $x^i_r, x^i_b$ be respectively the number of red and blue points that are removed from $T(v_i)$. The goal is to minimize $\sum_{i=1}^{\gamma^d} x^i_r + x^i_b$ such that the following conditions hold:
\begin{enumerate}
	\item for each $i\in [\gamma^d]$, $(N^i_r - x^i_r, N^i_b - x^i_b)$ is ($r,b$)-balanced.
	\item $(\sum_{i\in[\gamma^d]} x^i_r, \sum_{i\in[\gamma^d]} x^i_b)$ is ($r,b$)-balanced.
\end{enumerate}
\end{question}

\paragraph{Step 2.} After computing $\set{x^i_r, x^i_b}_{i\in[\gamma^d]}$, for each $i\in [\gamma^d]$, remove an {\em arbitrary} set of $x^i_r$ red and $x^i_b$ blue points from $T(v_i)$ and add them to $P_v$. Then, output an arbitrary $(r,b)$-fairlet decomposition of points $P_v$ which is guaranteed to be $(r,b)$-balanced by Step 1. 

\paragraph{Step 3.} For each non-empty child of $v$, $v_i$ (for $i\in [\gamma^d]$), run $\Call{\fairletDec}{v_i, r, b}$ which is guaranteed to be $(r,b)$-balanced by Step 1. 

Here is the main guarantee of our approach in the first step (i.e., ($r,b$)-fairlet decomposition).
\begin{theorem}\label{thm:main-fairlet}
There exists an $O(d\cdot n\cdot \log_{\gamma} n)$ time algorithm that given a point set $P \subseteq \mathbb{R}^d$ and balance parameters $(r,b)$, computes an $(r,b)$-fairlet decomposition of $P$ with respect to $\cost_{\med}$ whose expected cost is within $O(d\cdot (r^8+b^8)\cdot \log n)$ factor of the optimal $(r,b)$-fairlet decomposition of $P$ in expectation. 
\end{theorem}

\paragraph{Phase 2: merging ($r,b$)-fairlets into $k$ clusters.}
In this phase, we essentially follow the same approach as 
\cite{chierichetti2017fair}.
 
\begin{algorithm}
\caption{\Call{\clusterFairlet}{$Q$}: the algorithm returns an $(r,b)$-fair $k$-median of $P$ given an ($r,b$)-fairlet decomposition $Q$ of $P$}\label{alg:fairlet-to-clusters}
\begin{algorithmic}[1]
	\FORALL{fairlet $q_i \in Q$}
		\STATE {\bf let} an arbitrary point $c_i \in q_i$ be the center of $q_i$
		\STATE {\bf add} $|q_i|$ copies of $c_i$ to $\bar{P}$
	\ENDFOR
	\STATE $\sC:=\set{C_1, \cdots, C_k} \leftarrow \beta$-approximate $k$-median clustering of $\bar{P}$
	\STATE $\sC^* \leftarrow \set{\bigcup_{j: c_j \in C_i} q_j}_{i=1}^k$ \COMMENT{each fairlet joins the cluster of its center in $\sC$}
	\RETURN $\sC^*$
\end{algorithmic}
\end{algorithm}

\begin{theorem}[Fairlet to Fair Clustering]\label{thm:fairlet-to-clustering}
Suppose that $Q$ is an $\alpha$-approximate $(r,b)$-fairlet decomposition of $P$. Then, \Call{\clusterFairlet}{$Q$} returns an $(\alpha + (r+b)\cdot\beta)$-approximate $(r,b)$-fair $k$-median clustering of $P$ where $\beta$ denotes the approximation guarantee of the $k$-median algorithm invoked in \Call{\clusterFairlet}.
\end{theorem}

Finally, Theorem~\ref{thm:main-fairlet} and~\ref{thm:fairlet-to-clustering} together imply Theorem~\ref{thm:main-fair-k-median}.

%% file: fair_clustering.tex
\section{Fairlet Decomposition: a Top-down Approach on $\gamma$-HST}\label{sec:log-approx}

In this section, we provide a complete description of the first phase in our $(r,b)$-fair $k$-median algorithm (described in Section~\ref{sec:high-level}), namely our scalable ($r,b$)-fairlet decomposition algorithm. 

The first step in our algorithm is to embed the input point set into a $\gamma$-HST (for a value of $\gamma$ to be determined later in this section).
Once we build a $\gamma$-HST embedding $T$ of the input points $P\subseteq \mathbb{R}^d$, the question is how to partition the points into $(r,b)$-fairlets. 
We assume that each node $v\in T$ is augmented with extra information $N_r$ and $N_b$ respectively denoting the number of {\em red} and {\em blue} points in the subtree $T(v)$ rooted at $v$. 

To compute the total cost of a fairlet decomposition, it is important to specify the clustering cost model (e.g., $k$-median, $k$-center). Here, we define $\cost_{\med}$ to denote the cost of a fairlet (or cluster) with respect to the cost function of $k$-median clustering: for any subset of points $S\subset \mathbb{R}^d$, 
$\cost_{\med}(S) := \min_{p\in S}\sum_{q\in S} d(p,q)$ where $d(p,q)$ denotes the distance of $p,q$ in $T$.

In this section, we design a fast fairlet decomposition algorithm with respect to $\cost_{\med}$.
\begin{theorem}\label{thm:median-main}
There exists an $\tldO(n)$ time algorithm that given an $O(r^5 + b^5)$-HST embedding $T$ of the point set $P$ and balance parameters $(r,b)$, computes an $O(r^3+b^3)$-approximate $(r,b)$-fairlet decomposition of $P$ with respect to $\cost_{\med}$ on $T$. 
\end{theorem}
Thus, by Theorem~\ref{thm:median-main} and the bound on the expected distortion of embeddings into HST metrics (Theorem~\ref{thm:tree-metric-distortion}), we can prove Theorem~\ref{thm:main-fairlet}.
\begin{proofof}{\bf Theorem~\ref{thm:main-fairlet}.}
We first embed the points into an $O(r^5 + b^5)$-HST $T$ and then perform the algorithm guaranteed in Theorem~\ref{thm:median-main}. By Theorem~\ref{thm:tree-metric-distortion}, the expected distortion of our embedding to $T$ is $O(d\cdot \gamma \cdot \log_{\gamma} n)$ and by Theorem~\ref{thm:median-main}, there exists an algorithm that computes an $O(r^3+b^3)$-approximate fairlet-decomposition of $P$ with respect to distances in $T$. Hence, the overall algorithm achieves $O(d\cdot (r^8+b^8)\cdot \log n)$-approximation. 

Since the embedding $T$ can be constructed in time $O(d\cdot n \cdot \log n)$ and the fairlet-decomposition algorithm of Theorem~\ref{thm:median-main} runs in near-linear time, the overall algorithm also runs in $\tldO(n)$.   
\end{proofof}

Before describing the algorithm promised in Theorem~\ref{thm:median-main}, we define a modified cost function $\cost_{\smed}$ which is a simplified variant of $\cost_{\med}$ and is particularly useful for computing the cost over trees. Consider a $\gamma$-HST embedding of $P$ denoted by $T$ and assume that $\script{X}$ is a fairlet decomposition of $P$. 
Moreover, $h(D)$ denotes the height of $\lca(D)$ in $T$. For each fairlet $D$, $\cost_{\smed}(D)$ is defined as
\begin{align}\label{eq:fairlet-cost}
\cost_{\smed}(D) &:= \sum_{q\in D} d(\lca(D),q) = \Theta(\card{D} \cdot \gamma^{h(\lca(D))}).
\end{align}
In particular, $\cost_{\smed}(D)$ relaxes the task of finding ``the best center in fairlets'' and at the same time, its value is within a small factor of $\cost_{\med}(D)$.
\begin{claim}\label{clm:simple-median-cost}
Suppose that $T$ is a $\gamma$-HST embedding of the set of points $P$. For any $(r,b)$-fairlet $S$ of $P$, $\cost_{\med}(S) \leq \cost_{\smed}(S) \leq (r+b)\cdot \cost_{\med}(S)$ where the distance function is defined w.r.t. $T$.
\end{claim}

In the rest of this section we prove the following result which together with Claim~\ref{clm:simple-median-cost} imply Theorem~\ref{thm:median-main}.
\begin{lemma}\label{lem:simple-median-guarantee}
There exists a near-linear time algorithm that given an $O(r^5+b^5)$-HST embedding $T$ of the point set $P$ and balance parameters $(r,b)$, computes an $O(r^2+b^2)$-approximate $(r,b)$-fairlet decomposition of $P$ with respect to $\cost_{\smed}$ on $T$. 
\end{lemma}


\begin{lemma}\label{lem:constant-heavy-points}
For any tree $T$ with the root vertex $v$, the number of heavy points with respect to $v$ in the ($r,b$)-fairlet decomposition constructed by $\Call{\minHeavy}{v,r,b}$ is at most $O(r^2 + b^2)$ times the minimum number of heavy points in any valid ($r,b$)-fairlet decomposition of $T$. 
\end{lemma}

\subsection{Description of Step 1: Minimizing the Number of Heavy Points}\label{sec:heavy-points}
In this section, we show that \Call{\minHeavy} algorithm invoked by \Call{\fairletDec} finds an $O(r^2 + b^2)$-approximate solution of {\em Minimum Heavy Points} problem. 

The high-level overview of \Call{\minHeavy} is as follows. For any subset of points $D\subseteq P$, we can compute in $O(1)$ what the maximal size $(r,b)$-balanced subset of $D$ is: w.l.o.g. suppose that $N_r \geq N_b$ and $r\geq b$. If $N_r \leq {r\over b}\cdot N_b$, the collection is $(r,b)$-balanced. Otherwise, it suffices to greedily pick maximal size $(r,b)$-fairlets (see procedure \Call{\unbalancedPoints} for the formal algorithm). This simple observation implies a lower bound on the size of any optimal solution of \emph{Heavy Points Minimization} with respect to $v$ and we use this value to bound the approximation guarantee of \Call{\minHeavy} algorithm.
\begin{claim}\label{clm:minimum-point-removal}
\Call{\unbalancedPoints}{$N_r, N_b, r,b$} correctly computes the minimum number of points that is required to be removed from $N_r \cup N_b$ so that the remaining points become $(r,b)$-balanced. Moreover, the solution returned by the procedure only removes points form a single color class.
\end{claim}   
For each $i\in [\gamma^d]$, let $(\tilde{x}^i_r,\tilde{x}^i_b)$ be the output of \Call{\unbalancedPoints}{$N^i_r, N^i_b, r,b$}. 
\begin{corollary}\label{cor:trivial-lb}
Any $(r,b)$-fairlet decomposition of the points in $T(v)$ has at least $\sum_{i\in[\gamma^d]} \tilde{x}^i_r + \tilde{x}^i_b$ heavy points.
\end{corollary}

\begin{algorithm}[t]
\caption{\Call{\minHeavy}{$\set{N_r^i, N_b^i}_{i\in [\gamma^d]}, r, b, \gamma$}}
\begin{algorithmic}[1]
	\item[]
	\COMMENT{Stage $1$: lower bound on the number of heavy points} 
	\FORALL{non-empty children $i\in[\gamma^d]$ of $v$}
		\STATE $(x^i_r, x^i_b) \leftarrow \Call{\unbalancedPoints}{N^i_r, N^i_b, r, b}$ 
	\ENDFOR
	\STATE $y_r \leftarrow \sum_{i\in[\gamma^d]} x^i_r, y_b \leftarrow \sum_{i\in[\gamma^d]} x^i_b$
	\STATE $c_{\dom} \leftarrow \argmax_{c\in\set{r,b}} y_c, \bar{c_{\dom}} \leftarrow \set{r,b}\setminus c_{\dom}$
	
	\item[]
	\COMMENT{Stage $2$: add free points with color $\bar{c_\dom}$}
	\IF{$(\sum_{i\in[\gamma^d]} x^i_r, \sum_{i\in[\gamma^d]}x^i_b)$ is $(r,b)$-balanced}
		\STATE{break}
	\ENDIF
	\FORALL{non-empty children $i\in[\gamma^d]$ of $v$}
		\STATE $x^i_{\bar{c_{\dom}}} \leftarrow x^i_{\bar{c_\dom}} + \max(\Call{\extraPoints}{\bar{c_{\dom}}, N^i_r - x^i_r, N^i_b - x^i_b, r, b}, y_{c_{\dom}} - y_{\bar{c_{\dom}}})$	
	\ENDFOR

	\item[]
	\COMMENT{Stage $3$: add points of non-saturated $(r,b)$-fairlets}
	\IF{$(\sum_{i\in[\gamma^d]} x^i_r, \sum_{i\in[\gamma^d]}x^i_b)$ is $(r,b)$-balanced}
		\STATE{break} 
	\ENDIF

		\FORALL{non-empty children $i\in[\gamma^d]$ of $v$}
		\STATE $(n_r, n_b) \leftarrow \Call{\freeFairlet}{N^i_r - x^i_r, N^i_b - x^i_b, r, b}$ 
		\STATE $x^i_r \leftarrow x^i_r + n_r$, $x^i_b \leftarrow x^i_b + n_b$
	\ENDFOR	
	\RETURN $(\set{x^i_r, x^i_b}_{i\in[\gamma^d]})$
\end{algorithmic}
\end{algorithm}

\paragraph{Stage 1: minimum number of heavy points.}
If $\sum_{i\in[\gamma^d]} \tilde{x}^i_r$ red points together with $\sum_{i\in[\gamma^d]}\tilde{x}^i_b$ blue points form an $(r,b)$-balanced collection, then \Call{\minHeavy} technically terminates at the end of stage $1$ and the solution returned by \Call{\minHeavy} achieves the minimum possible number of heavy points. 
However, in general, the collection with $\sum_{i\in[\gamma^d]} \tilde{x}^i_r$ red points and $\sum_{i\in[\gamma^d]}\tilde{x}^i_b$ may not form an $(r,b)$-balanced collection. Next, we show that we can always pick at most $rb(\sum_{i\in[\gamma^d]} \tilde{x}^i_r + \tilde{x}^i_b)$ additional heavy points and keep both all subtrees rooted at children of $v$ and the set of heavy points $(r,b)$-balanced.

\begin{algorithm}
\caption{\Call{\unbalancedPoints}{$N_r, N_b, r, b$}: returns the minimum number of points that are required to be removed so that $(N_r, N_b)$ become $(r,b)$-balanced.}\label{alg:min-unbalanced}
\begin{algorithmic}[1]
	\IF{$N_r \geq N_b$} \RETURN {$(N_r - \floor{N_b \cdot {r\over b}}, 0)$}
	\ENDIF
	\RETURN $(0, N_b - \floor{N_r \cdot {r\over b}})$
\end{algorithmic}
\end{algorithm}

Another structure we will refer to in the rest of this section is {\em saturated ($r,b$)-fairlets}. A fairlet $D$ is a saturated $(r,b)$-fairlet if it has {\em exactly} $r+b$ points; $r$ points from color $c$ and $b$ points from color $\bar{c}$. 
\paragraph{Stage 2: Adding free points.}
If the ``must-have'' heavy points are not $(r,b)$-balanced, then one color is {\em dominant}. For a color class $c\in\set{r,b}$, a collection of points $S$ is $c$-dominant if $\card{S_c} \geq {r\over b}\cdot \card{S_{\bar{c}}}$. Moreover, the collection is {\em minimally-balanced} $c$-dominant if $S$ is $(r,b)$-balanced but it will be no longer $(r,b)$-balanced even if we remove a single point of color $\bar{c}$.

Let $c$ be the dominant color in the {heavy points}. Then, we inspect all children of $v$ and if there exits a child in which $\bar{c}$ is dominant, we borrow as many points of color $\bar{c}$ as we can (we need to keep the subtree $(r,b)$-balanced, see \Call{\extraPoints} procedure) till either the set of heavy points becomes $(r,b)$-balanced or all subtrees rooted at children of $v$ become minimally-balanced $c$-dominant. It is straightforward to show that at most ${b\over r}\cdot \card{S_c}$ points of color $\bar{c}$ will be borrowed from the children of $v$ in this phase.

\begin{algorithm}
\caption{\Call{\extraPoints}{$c, N_r, N_b, r, b$}: returns the maximum number of points of color $c$ that can be removed from the set $(N_r, N_b)$ such that they remain $(r,b)$-balanced.}\label{alg:extra-points}
\begin{algorithmic}[1]
	\IF{$N_c \leq N_{\bar{c}}$} \RETURN $0$ \ENDIF
	\RETURN $\ceil{N_{\bar{c}} \cdot {b\over r}}$
\end{algorithmic}
\end{algorithm}

\begin{lemma}\label{lem:structure-phase-2}
Suppose that the set of heavy points is $c$-dominant. If the set of heavy points is not $(r,b)$-balanced at the end of stage 2, then for each $i\in [\gamma^d]$, the set of points in the subtree rooted at $v_i$ is minimally-balanced $c$-dominant.
\end{lemma}
\begin{corollary}\label{cor:non-maximal}
Suppose that the set of heavy points is $c$-dominant. If the set of heavy points is not $(r,b)$-balanced at the end of stage 2, then for each $i\in [\gamma^d]$, the set of points in the subtree rooted at $v_i$ have an $(r,b)$-fairlet decomposition with at most one {non-saturated} $(r,b)$-fairlet. 
\end{corollary}

\paragraph{Stage 3: Non-saturated fairlets.}
Here, we show that we can increase the number of heavy points by at most a factor of $O(rb)$ 
and make both the set of heavy points and the set of points in all subtrees rooted at children of $v$ $(r,b)$-balanced. Let $ns$ denote the total number of non-saturated fairlets in the subtree rooted at $v$. We consider two cases depending on the value of $ns$ and the total number of heavy points $N_\heavy$ that do not belong to any saturated fairlets (in particular, $\card{N_\heavy} \leq \sum_{i\in[\gamma^d]} \tilde{x}^i_r + \tilde{x}^i_b$): 

\paragraph{Case 1: $\boldsymbol{ns \leq b\cdot N_\heavy}$.}
If we add all non-saturated fairlets, since the rest of fairlets in the subtree rooted at children of $v$ are saturated $(r,b)$-balanced, then this ``extended'' collection of heavy points has to be $(r,b)$-balanced. Otherwise, the whole data set itself is not $(r,b)$-balanced which is a contradiction. Moreover, the total number of heavy points is at most $ns \times (r+b) = O(rb\cdot N_\heavy)$

\paragraph{Case 2: $\boldsymbol{ns > b\cdot N_\heavy}$.} 
Here we show that after adding at most $b \cdot N_{\heavy}$ non-saturated $(r,b)$-fairlets, the set of heavy points becomes $(r,b)$-balanced. Let $r_i$ and $b_i$ ($r_i\geq b_i$) specify the size of the non-saturated fairlet that belongs to the $i$-th child of $v$. Note that since all fairlets are $c$-dominant, $r_i$ denotes the number of points of color $c$.  

Moreover, in any  non-saturated $(r,b)$-fairlet, ${r_i \over b_i} < {r \over b}$, which implies that $r_i b \leq rb_i -1$. 
Let $Q$ denote the set of children of $v$ whose non-saturated fairlets are picked. After adding all points in these non-saturated fairlets, 
\begin{align}
\# \text{points of color } c 
&\leq N_{\heavy} + \sum_{j\in Q} {r_j} \leq N_{\heavy} + \sum_{j\in Q} ({r \over b}\cdot  b_j - {1\over b}) \leq {r\over b} \sum_{j\in Q} b_j \rhd\text{since $\card{Q} =b\cdot N_{\heavy}$},\label{eq:heavy-point-count-c}\\
\# \text{points of color } \bar{c} &\geq {r\over b} \sum_{j\in Q} b_j.\label{eq:heavy-point-count-bar-c}
\end{align}
Moreover, since in the beginning of the process the number of points of color $c$ is more than the number of points of color $\bar{c}$ and also in each non-saturated fairlest the number of points of color $c$ is more than the number of points of color $\bar{c}$, at the end of the process, in heavy points, the size of color $c$ is larger than the size of color $\bar{c}$.
Thus, by~\eqref{eq:heavy-point-count-c}~and~\eqref{eq:heavy-point-count-bar-c}, at the end of stage $3$, the extended heavy points has size $O(rb\cdot N_{\heavy})$ and is $(r,b)$-balanced as promised in Lemma~\ref{lem:constant-heavy-points}. 

\paragraph{Runtime analysis of \minHeavy.} Here we analyze the runtime of \Call{\minHeavy} which corresponds to step $1$ in \Call{\fairletDec}. Note that stage $1$ only requires $O(1)$ operations on the number of red and blue points in $T(v)$. Each of stage $2$ and stage $3$ requires $O(1)$ operations on the number of red and blue points in all non-empty children of $T(v)$. Although the number of children of $T(v)$ can be as large as $\gamma^d$, for each node $v$ in $T$, \Call{\minHeavy} performs $O(1)$ operations on the number of red and blue points in $T(v)$ exactly twice: when it is called on $v$ and the parent of $v$. Hence, in total $\Call{\minHeavy}$ performs $O(1)$ time on each node in $T$ which in total is $O(n)$.

\begin{algorithm}
\caption{\Call{\freeFairlet}{$N_r, N_b, r, b$}: returns the non-saturated fairlet in a set with $(N_r, N_b)$ points.}\label{alg:non-saturated-fairlet}
\begin{algorithmic}[1]
	\IF{$N_r \leq N_{b}$} 
		\STATE $z_r \leftarrow N_r - \floor{N_r \over b}\cdot b$, $z_b \leftarrow N_b - \floor{N_r \over b}\cdot r$
	\ELSE{} 
		\STATE $z_b \leftarrow N_b - \floor{N_b \over b}\cdot b$, $z_r \leftarrow N_r - \floor{N_b \over b}\cdot r$
	\ENDIF
	\RETURN $(z_r, z_b)$
\end{algorithmic}
\end{algorithm}

%% file: experiment.tex
\section{Experiments}\label{sec:experiment}
\begin{table*}[!t]\label{tbl:comparison}
\centering
\begin{minipage}{\textwidth}
\centering
\resizebox{\textwidth}{!}{%
\renewcommand{\arraystretch}{1.2}
\begin{tabular}{|c|c||c|c||c|c|}
\hline 
 \multirow{ 2}{*}{\bf Dataset} & \multirow{ 2}{*}{\bf Balance} & \multicolumn{2}{c||} {\bf Fairlet Decomposition Cost}   & \multicolumn{2}{c|}{{\bf Fair Clustering Cost} ($k=20$)} \\
 \cline{3-6}
 &	& \citep{chierichetti2017fair} &  {\bf Ours} & \citep{chierichetti2017fair} &  {\bf Ours} \\
\hline
Diabetes (1000 points) & $0.8$\footnote{In~\citep{chierichetti2017fair}, based on the description of the experiment setup, the desired balance in all three datasets (including Diabetes) are 0.5. However, for Diabetes dataset, they have achieved the higher balance of value 0.8.} & $\sim9836$ & $2971$ & $\sim 9909$& 4149\\ 
\hline
 Bank (1000 points) & $0.5$ & $\sim5.46 \times 10^5$ & $5.24 \times 10^5$  & $\sim5.55\times 10^5$ & $6.03\times 10^5$\\
\hline
 Census (600 points) & $0.5$ & $\sim3.59 \times 10^7$ & $2.31\times 10^7$ & $\sim3.65 \times 10^7$ & $2.41\times 10^7$\\
 \hline
\end{tabular}
}
\caption{The table compares the performance of our fairlet-decomposition algorithm and the algorithm of~\citep{chierichetti2017fair}. We remark that the number for~\citep{chierichetti2017fair} mentioned in this table are not explicitly stated in their paper and we have extracted them from Figure 3 in their paper. Note that the cost denotes the total distances of the points to their fairlet/cluster centroids.}
\label{table:results}
\end{minipage}
\end{table*}

\begin{figure*}[!h]
\centering
\subfigure{\includegraphics[width=0.49\textwidth]{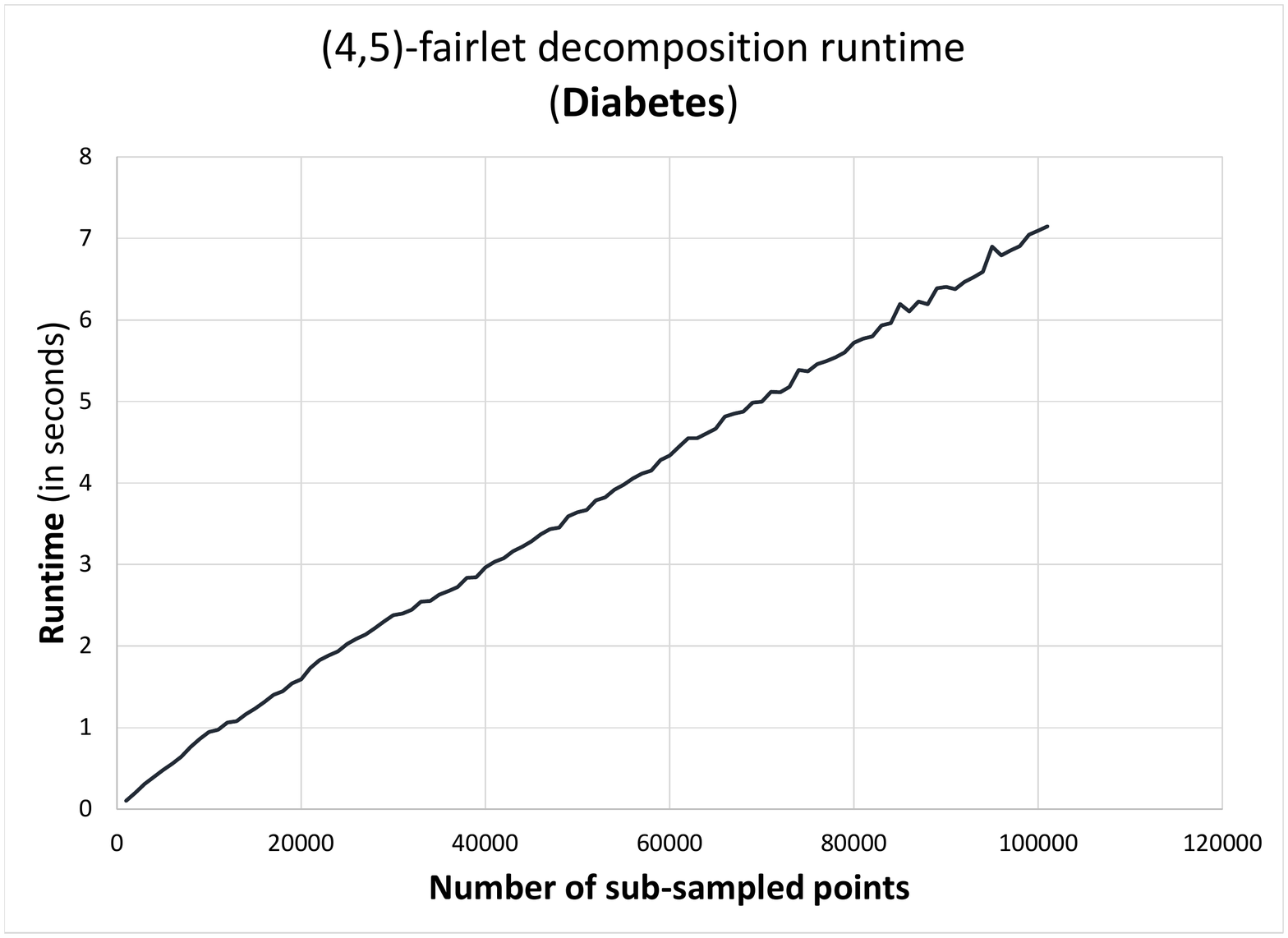}}
\subfigure{\includegraphics[width=0.49\textwidth]{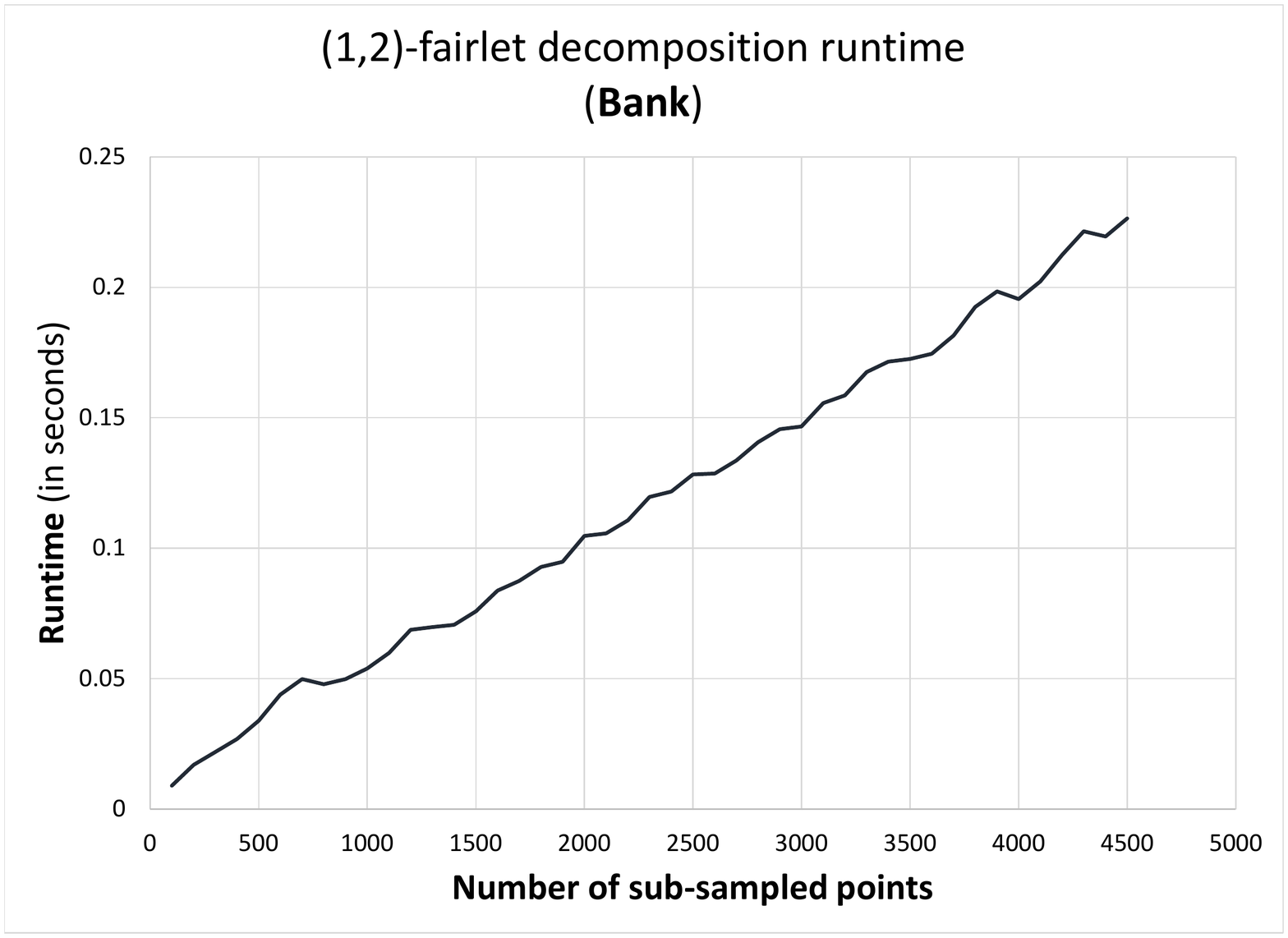}}
\centering
\subfigure{\includegraphics[width=0.49\textwidth]{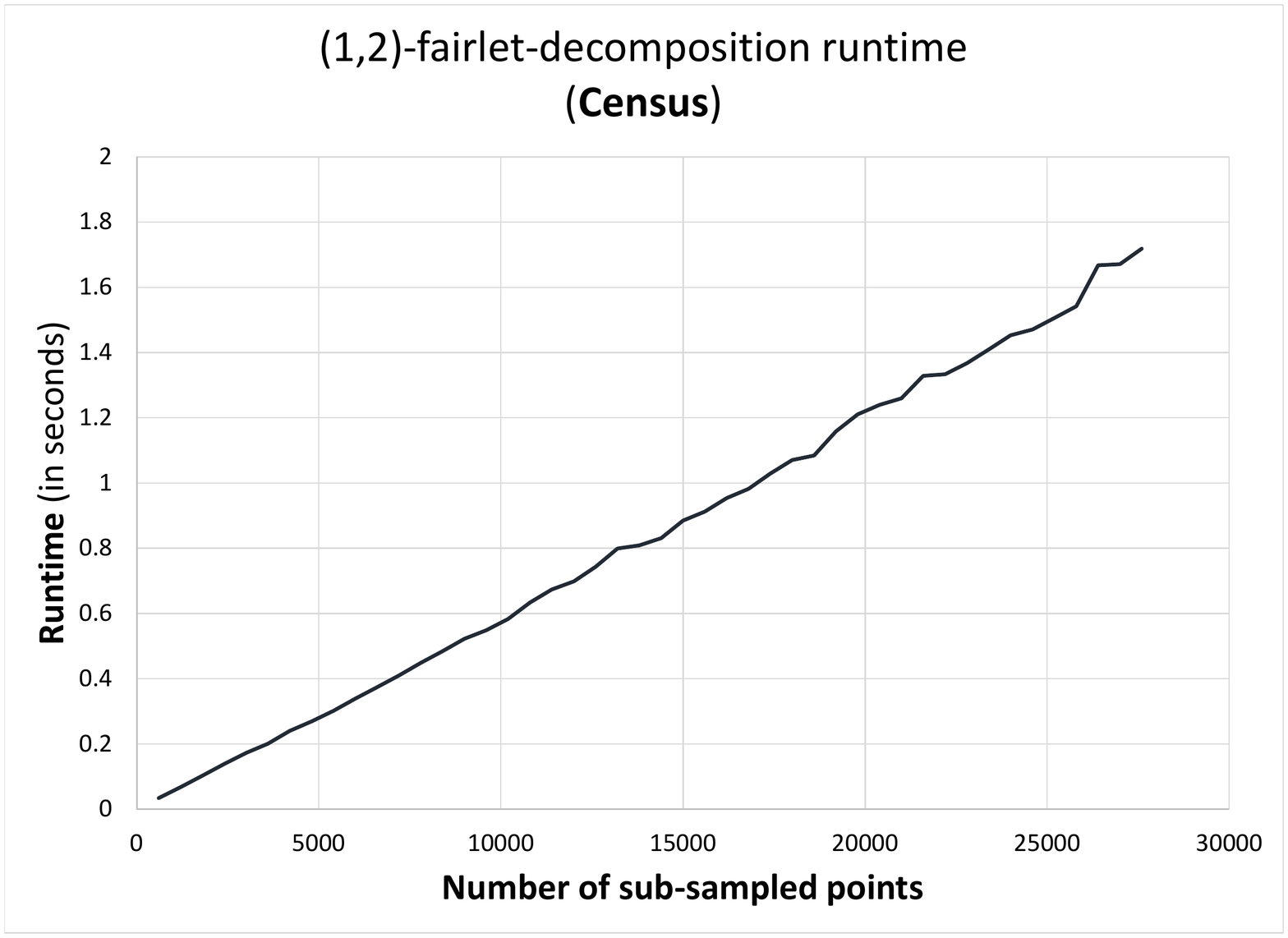}}
\subfigure{\includegraphics[width=0.49\textwidth]{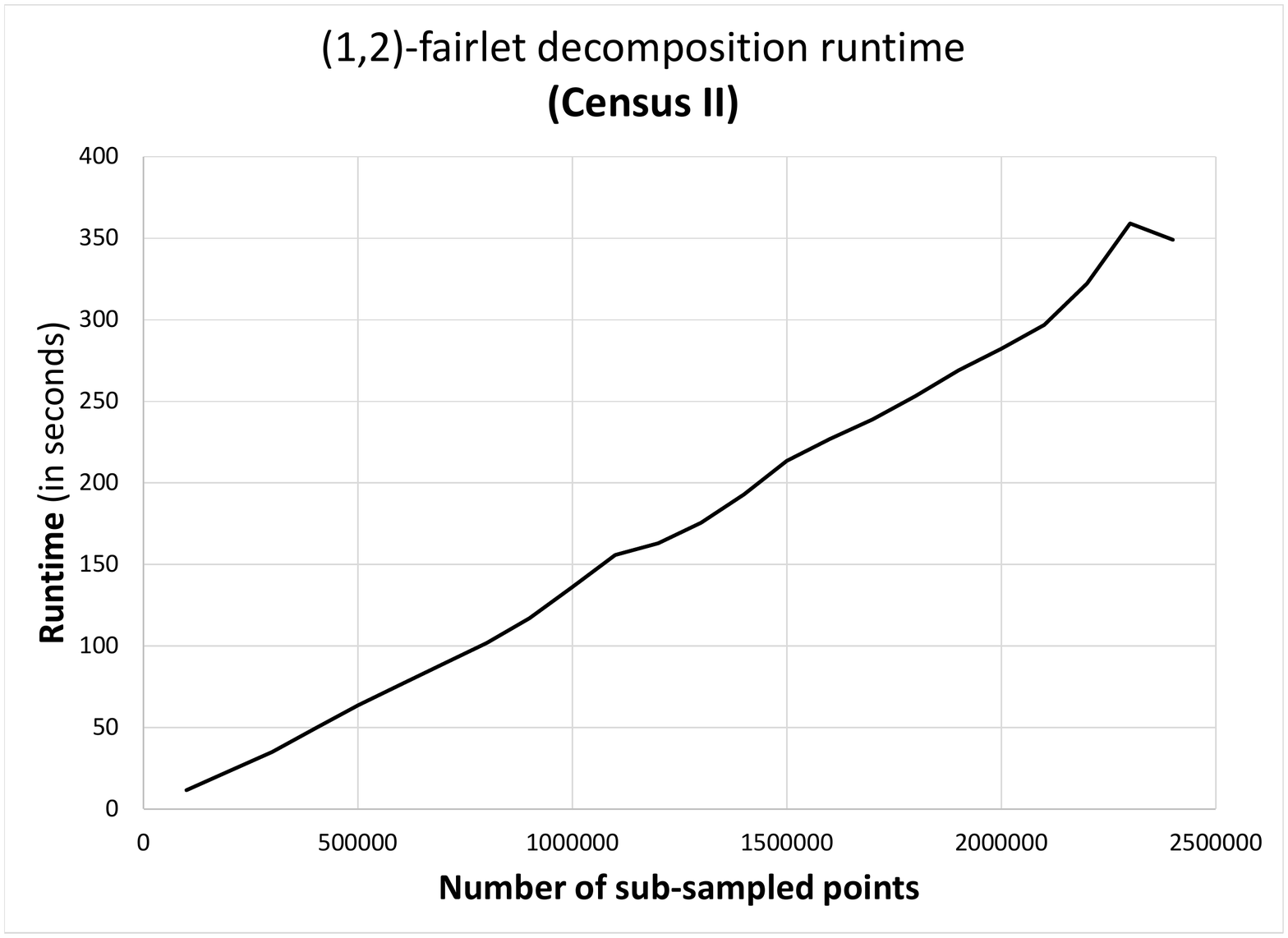}}
\caption{Each figure captures the running time of our fairlet decomposition algorithms with the specified balance parameter
on different number of sample points from one of the four datasets: Diabetes, Bank, Census and Census II.}\label{fig:runtime}
\end{figure*}

In this section we show the performance of our proposed algorithm for $(r,b)$-fair $k$-median problem on three different standard data sets considered in~\citep{chierichetti2017fair} which are from UCI Machine Learning Repository~\citep{Dua:2017}\footnote{\href{https://archive.ics.uci.edu/ml/datasets/diabetes}{https://archive.ics.uci.edu/ml/datasets/diabetes}}. Furthermore, to exhibit the performance of our algorithms on large and high-dimensional scale datasets, we consider an additional data set.

\begin{itemize}
\item{\bf Diabetes.} The dataset\footnote{\href{https://archive.ics.uci.edu/ml/datasets/diabetes+130-us+hospitals+for+years+1999-2008}{https://archive.ics.uci.edu/ml/datasets/diabetes+130-us+hospitals+for+years+1999-2008}} represents 10 years of clinical care at 130 US hospitals and in particular represent the information and outcome of patients pertaining to diabetes~\citep{strack2014impact}. Points are in $\mathbb{R}^2$ and dimensions correspond to two attributes (``age'', ``time-in-hospital''). Moreover, we consider ``gender'' as the sensitive attribute.  
\item{\bf Bank.} The dataset\footnote{\href{https://archive.ics.uci.edu/ml/datasets/Bank+Marketing}{https://archive.ics.uci.edu/ml/datasets/Bank+Marketing}} is extracted from marketing campaigns of a Portuguese banking institution~\citep{moro2014data}. Among the information about the clients, we selected (``age'', ``balance'', ``duration-of-account'') as attributes to represent the dimensions of the points in the space. Moreover, we consider ``marital-status'' as the sensitive information.   
\item{\bf Census.} The dataset\footnote{\href{https://archive.ics.uci.edu/ml/datasets/adult}{https://archive.ics.uci.edu/ml/datasets/adult}} contains the records extracted from 1994 US Census~\citep{kohavi1996scaling}. We picked attributes (``age'' , ``fnlwgt'', ``education-num'', ``capital-gain'', ``hours-per-week'') to represent the points in the space. Moreover, we consider ``gender'' as the sensitive attribute. 
\item{\bf Census II.} The dataset\footnote{\href{https://archive.ics.uci.edu/ml/datasets/US+Census+Data+(1990)}{https://archive.ics.uci.edu/ml/datasets/US+Census+Data+(1990)}} contains the records extracted from 1990 US Census. We picked 25  numeric attributes to represent points in the space. Moreover, we consider ``gender'' as the sensitive attribute. 
\end{itemize} 

\begin{table}[!h]
\centering
\resizebox{0.75\textwidth}{!}{%
\renewcommand{\arraystretch}{1}
\begin{tabular}{|c||c|c|c|}
\hline 
 {\bf Dataset} & {\bf Dimension} & {\bf Number of points} & {\bf Sensitive attribute}\\
\hline
\hline
Diabetes & $2$ & $101,765$ & gender\\ 
\hline
 Bank & $3$ & $4,520$ & marital-status\\
\hline
 Census & $5$ & $32,560$ & gender\\
 \hline
 Census II & $25$ & $2,458,285$ & gender\\
 \hline
\end{tabular}
}
\caption{The description of the three datasets used in our empirical evaluation. In each dataset, the goal is find a fair $k$-median with respect to the sensitive attribute.}
\label{table:dataset-description-1}
\end{table}

\paragraph{Algorithm.} We essentially implement the algorithm described in Section~\ref{sec:log-approx}.\footnote{Our code is publicly available at~\url{https://github.com/talwagner/fair_clustering}.} However, instead of building $\poly(r,b)$-HST, in our implementation, we embed the points into a $2$-HST. After computing a fairlet-decomposition of the points with given balance parameters, we run an existing $K$-medoids clustering subroutine\footnote{\href{https://www.mathworks.com/help/stats/kmedoids.html}{https://www.mathworks.com/help/stats/kmedoids.html}}.  

\paragraph{Results.} Comparing the cost of the solution returned by our fairlet decomposition algorithm with the result of~\citep{chierichetti2017fair} (as in Table~\ref{table:results}) shows that we achieve empirical improvements on all instances. The main reason is that our algorithm is particularly efficient when the input pointset lies in a low dimensional space which is the case in all three datasets ``Diabetes'', ``Bank'' and ``Census''.   
Moreover, unlike~\citep{chierichetti2017fair}, for each dataset, we can afford running our algorithm on the whole dataset (see Table~\ref{table:dataset-runtime}). Empirically, the running time of our algorithm scales almost linearly in the number points in the input pointset (see Figure~\ref{fig:runtime}). 

In Figure~\ref{fig:runtime} and both Table~\ref{table:results} and~\ref{table:dataset-runtime}, the reported runtime for each sample size $S$ is the median runtime of our algorithm on $10$ different sample sets from the given pointset each of size $S$. 
\begin{table}[!h]
\centering
\resizebox{0.65\textwidth}{!}{%
\renewcommand{\arraystretch}{1}
\begin{tabular}{|c||c||c|c|}
\hline 
 \multirow{ 2}{*}{\bf Dataset} & {\bf Target} &\multicolumn{2}{c|} {\bf Runtime for $k=20$ (in sec)}\\
 \cline{3-4}
 &  {\bf Balance} &\bf{Fairlet dec.} &  {\bf Total} \\
\hline
\hline
 Diabetes & $0.8$ & $7.42$ & $14$\\ 
\hline
 Bank & $0.5$ & $0.23$ & $7.63$ \\
\hline
 Census & $0.45$ & $5.18$ & $14.19$ \\
 \hline
 Census II & $0.5$ & $349.08$ & $750.09$ \\
 \hline
\end{tabular}
}
\caption{The performance of our algorithm on all points in each dataset. We provide the runtime of both fairlet decomposition and the whole clustering process. Since Census dataset is not $(1,2)$-balanced, we picked a lower balance-threshold for this dataset.}
\label{table:dataset-runtime}
\end{table}

%% file: missing_proofs.tex
\section{Missing Proofs}\label{sec:missing_proof}
\begin{proofof}{\bf Lemma~\ref{lem:simple-median-guarantee}.}
The proof is by induction on height of $v$ in $T$. The base case is when $v$ is a leaf node in $T$ and the algorithm trivially finds an optimal solution in this case. Suppose that the induction hypothesis holds for all vertices of $T$ at height $h-1$. Here, we show that the statement holds for the vertices of $T$ at height $h$ as well.

Let $\opt$ denote an optimal ($r,b$)-fairlet decomposition of the points in $T(v)$ with respect to $\cost_{\smed}$. 
Next, we decompose $\opt$ into $\gamma^d + 1$ parts: $\set{\opt_i}_{i\in [\gamma^d]}$ and $\opt_{\heavy}$. 
For each $i\in[\gamma^d]$, $\opt_i$ denotes the set of fairlets in $\opt$ whose $\lca$ are in $T(v_i)$. Moreover, $\opt_{\heavy}$ denotes the set of heavy fairlets with respect to $v$ and $\heavy_{\opt}$ denotes the set of heavy points with respect to $v$ in $\opt$. Lastly, $\heavy^i_{\opt} := \heavy_\opt \cap T(v_i)$ denotes the set of heavy points with respect to $v$ in $\opt$ that are contained in $T(v_i)$. 
 
Let $\sol$ denote the solution returned by \Call{\fairletDec}{$v, r, b$}. Similarly, we decompose $\sol$ into $\gamma^d+1$ parts:  $\set{\sol_i}_{i\in [\gamma^d]}$ and $\sol_{\heavy}$. Moreover, $\heavy_{\sol}$ denotes the set of heavy points with respect to $v$ in $\sol$ and for each $i\in [\gamma^d]$, $\heavy^i_{\sol} := \heavy_\sol \cap T(v_i)$ denotes the set of heavy points with respect to $v$ in $\sol$ that are contained in $T(v_i)$.

\begin{claim}\label{clm:aug-cost}
For each $i\in [\gamma^d]$, there exists an $(r,b)$-fairlet decomposition of $P_i\setminus \heavy_{\sol}^i$ of cost at most $\cost(\opt_i) + (|\heavy^i _\opt| + |\heavy^i_\sol| \cdot (r+b)) \cdot (r^2 + b^2) \cdot \gamma^{h-1}$ where $P_i$ is the set of points contained in $T(v_i)$.
\end{claim}
Hence, by the induction hypothesis, for each $i\in [\gamma^d]$,
\begin{align}\label{eq:children-cost}	
\cost(\sol_i) \leq  c\cdot (r^2 + b^2) \cdot (\cost(\opt_i) + (|\heavy^i _\opt| + |\heavy^i_\sol| \cdot (r+b)) \cdot (r^2 + b^2) \cdot \gamma^{h-1}).
\end{align}
Next, we bound the cost of $\sol$ by Lemma~\ref{lem:constant-heavy-points} and~\eqref{eq:children-cost} as follows: 
\begin{align*}
\cost(\sol) &= \cost(\sol_{\heavy}) + \sum_{i\in[\gamma^d]} \cost(\sol_{i}) \\
			 &\leq \eta_{\heavy}\cdot (r^2 + b^2) \cdot \cost(\opt_{\heavy}) \\ 
			 &\;\;\; + c \cdot (r^2+b^2) \cdot ({\eta_{\heavy} (r+b)^5\over \gamma} \cdot \cost(\opt_{\heavy}) + \sum_{i\in [k^d]}\cost(\opt_i))\\
			 &\leq c\cdot (r^2 + b^2) \cdot \cost(\opt) \\
			 &\;\;\; + (\eta_\heavy - {c\over 2})\cdot (r^2+ b^2) \cdot \cost(\opt_{\heavy})\rhd \text{By setting } \gamma := 2\eta_{\heavy}(r+b)^5\\
			 &\leq c\cdot (r^2 + b^2) \cdot \cost(\opt) \rhd c \geq 2\eta_\heavy
\end{align*}
\end{proofof}

\begin{proofof}{\bf Claim~\ref{clm:aug-cost}.}
Consider the fairlet decomposition $\opt_i$ on $P_i\setminus \heavy_{\opt}^i$. A fairlet $D\in \opt_i$ is \emph{affected} if it contains a point $p \in \heavy_{\sol}^i$. 

We define the set of \emph{affected points} as $\bar{P}_i = \heavy^i_{\opt} \cup \bigcup_{D\in \bar{\opt}_i} D$ to denote the union of the points in the affected fairlets (i.e., $\bigcup_{D\in \bar{\opt}_i} D$) and the set $\heavy_\opt^i$ (whose points do not belong to any of fairlets in $\opt_i$).

Next, we bound the cost of the fairlet decomposition which is constructed by augmenting the set of fairlets $\opt_i \setminus \bar{\opt}_i$ with the set of affected points $\bar{P}_i$. 

Let $Q_0$ denote the set of affected points $\bar{P}_i$. We augment the fairlet decomposition in three steps:

\paragraph{Step 1.} In this step, we create as many $(r,b)$-balanced fairlets using the affected points $Q_0$ only. Note that the contribution of each point involved in such fairlets is $h_T(v_i)$ where $h_T(v_i)$ denotes the distance of $v_i$ from the leaves in $T(v_i)$. Let $Q_1 \subseteq Q_0$ denote the set of affected points that do not join any fairlets at the end of this step. Note that all points in $Q_1$ are of the same color $c$.  
\paragraph{Step 2:} Next, we add as many points of $Q_1$ as possible to the existing fairlets in $\opt_i\setminus \bar{\opt}_i$ while preserving the $(r,b)$-balanced property. Now the extra cost incurred by each points of $Q_1$ that joins a fairlet in this step is at most $(r+b)\cdot h_T(v_i)$. Let $Q_2\subset Q_1$ be the set of points that do not belong to an fairlets by the end of the second phase. Note that at the end of this step, if $Q_2$ is non-empty, then all fairlets are maximally-balanced $c$-dominant (a fairlet $S$ is maximally-balanced $c$-dominant if (1) in $S$, the number of points of color $c$ are larger than the number of points in color $\bar{c}$, (2) the set $S$ is $(r,b)$-balanced, and (3) adding a point of color $c$ to $S$ makes it unbalanced).  
\paragraph{Step 3:} Finally, we show that by mixing the points of at most $b\cdot|Q_2|$ existing fairlets with the set $Q_2$, we can find an $(r,b)$-balanced fairlet decomposition of the involved points and the contribution of each such point to the total cost is at most $h_T(v_i)$. Note that since the set of all points we are considering is $(r,b)$-balanced, not all of the so far constructed fairlets are saturated (i.e., has size exactly $r+b$). In particular, we show that there exists a set of non-saturated fairlets $\mathcal{X}$ of size at most $b\cdot |Q_2|$ whose addition to $Q_2$ constitutes a $(r,b)$-balanced set. For each fairlet $D\in \mathcal{X}$, 
\begin{align*}
\quad |c_{D}| < {r\over b} |\bar{c}_D|\Rightarrow b\cdot |c_D| \leq r\cdot |\bar{c}_D| - 1,
\end{align*}
where $c_D$ and $\bar{c}_D$ respectively denotes the set of points of color $c$ and $\bar{c}$ in $D$.
This implies that after picking at most $|Q_2|$ non-saturated fairlets (i.e., the fairlets in $\mathcal{X}$),
\begin{align*}
b\cdot |c_{\mathcal{X}}| \leq r\cdot |\bar{c}_{\mathcal{X}}| - b\cdot |Q_2| \Rightarrow b\cdot (|c_{\mathcal{X}}| + |Q_2|) \leq r\cdot |\bar{c}_{\mathcal{X}}|,
\end{align*}
where $c_{\mathcal{X}}$ and $\bar{c}_{\mathcal{X}}$ respectively denotes the set of points of color $c$ and $\bar{c}$ in $\bigcup_{D\in \mathcal{X}}D$.
Hence, the set of points $Q_2 \cup \bigcup_{D\in \mathcal{X}} D$ is $(r,b)$-balanced. Moreover, the cost of this step is at most $|Q_2|\cdot b \cdot (r+b) \cdot h_T(v_i)$. 

Altogether, there exists a fairlet decomposition of $P_i\setminus \heavy_{\sol}^i$ of cost at most 

\begin{align*}
&\quad\; \cost(\opt_i) + |Q_0\setminus Q_1| \cdot h_T(v_i) + |Q_1\setminus Q_2|\cdot (r+b) \cdot h_T(v_i) + |Q_2| \cdot b \cdot (r+b) \cdot h_T(v_i) \\
&\leq \cost(\opt_i) + |Q_0| \cdot b \cdot (r+b) \cdot h_T(v_i) \\ 
&\leq \cost(\opt_i) + (|\heavy^i _\opt| + |\heavy^i_\sol| \cdot (r+b)) \cdot (r^2 + b^2) \cdot \gamma^{h-1}
\end{align*}
\end{proofof}

\begin{proofof}{\bf Theorem~\ref{thm:fairlet-to-clustering}}
For a pointset $X$, let $\opt_{k\text{-fair}}(X)$ and $\opt_{\text{fairlet}}(X)$ respectively denote an optimal $(r,b)$-fair $k$-median and an optimal $(r,b)$-fairlet decomposition of $X$. It is straightforward to see that for any set of point $X$, $\cost(\opt_{\text{fairlet}}(X)) \leq \cost(\opt_{k\text{-fair}}(X))$ and in particular,
\begin{align}\label{eq:fairlet-to-k-median}
\cost(Q) \leq \alpha \cdot \cost(\opt_{k\text{-fair}}(P)).
\end{align}
Let $N$ denote the set of the centers of fairlets in $Q$. For a set of points $X$, let $\opt_{k\text{-median}}(X)$ denotes an optimal $k$-median clustering of $X$ (note that there is not fairness requirement). Since $C\subseteq P$, the optimal $k$-median cost of $N$ is smaller than the optimal $k$-median cost of $P$. Since $\bar{P}$ contains at most $(r+b)$ copies of each point of $N$, by assigning all copies of each point $p\in N$ in $\bar{P}$ to the center of $p$ in an optimal $k$-median clustering of $N$,   
\begin{align}
\cost(\opt_{k\text{-median}}(\bar{P})) &\leq (r+b) \cdot \cost(\opt_{k\text{-median}}({N})) \leq (r+b) \cdot \cost(\opt_{k\text{-median}}({P})).\label{eq:clustering-approx}
\end{align}
As \Call{\clusterFairlet}\ returns a $\beta$-approximate $k$-median clustering of $\bar{P}$, and by~\eqref{eq:fairlet-to-k-median}-\eqref{eq:clustering-approx}, the cost of the clustering $\sC$ constructed by \Call{\clusterFairlet}\ is  

Since the distance of each point $p_i \in P$ to the center of its cluster in $\sC^*$ is less than the sum of its distance to the center of its fairlet $c_i$ in $Q$ and the distance of $c_i$ to its center in $\sC$, we can bound the cost of $\sC^*$ in terms of the costs of $\sC$ and $Q$ as follows:
\begin{align*}
\cost(\sC^*) &\leq \cost(Q) + \cost(\sC)\\
&\leq \alpha \cdot \cost(\opt_{k\text{-fair}}(P)) &&\rhd \text{By~\eqref{eq:fairlet-to-k-median}} \\ 
&\quad +\beta \cdot \cost(\opt_{k\text{-median}}(\bar{P}))\\
&\leq \alpha \cdot \cost(\opt_{k\text{-fair}}(P)) \\ 
&\quad + \beta \cdot (r+b) \cdot \cost(\opt_{k\text{-median}}({P})) &&\rhd \text{By~\eqref{eq:clustering-approx}} \\
&\leq (\alpha + \beta\cdot (r+b)) \cdot  \cost(\opt_{k\text{-fair}}(P))
\end{align*}
\end{proofof}

%% file: main.bbl
\begin{thebibliography}{15}
\providecommand{\natexlab}[1]{#1}
\providecommand{\url}[1]{\texttt{#1}}
\expandafter\ifx\csname urlstyle\endcsname\relax
  \providecommand{\doi}[1]{doi: #1}\else
  \providecommand{\doi}{doi: \begingroup \urlstyle{rm}\Url}\fi

\bibitem[Bartal(1996)]{bartal1996probabilistic}
Y.~Bartal.
\newblock Probabilistic approximation of metric spaces and its algorithmic
  applications.
\newblock In \emph{Foundations of Computer Science, 1996. Proceedings., 37th
  Annual Symposium on}, pages 184--193. IEEE, 1996.

\bibitem[Bera et~al.(2019)Bera, Chakrabarty, and Negahbani]{fairclustering19}
S.~K. Bera, D.~Chakrabarty, and M.~Negahbani.
\newblock Fair algorithms for clustering.
\newblock \emph{arXiv preprint arXiv:1901.02393}, 2019.

\bibitem[Bercea et~al.(2018)Bercea, Gro{\ss}, Khuller, Kumar, R{\"{o}}sner,
  Schmidt, and Schmidt]{bercea2018cost}
I.~O. Bercea, M.~Gro{\ss}, S.~Khuller, A.~Kumar, C.~R{\"{o}}sner, D.~R.
  Schmidt, and M.~Schmidt.
\newblock On the cost of essentially fair clusterings.
\newblock \emph{CoRR}, abs/1811.10319, 2018.

\bibitem[Chierichetti et~al.(2017)Chierichetti, Kumar, Lattanzi, and
  Vassilvitskii]{chierichetti2017fair}
F.~Chierichetti, R.~Kumar, S.~Lattanzi, and S.~Vassilvitskii.
\newblock Fair clustering through fairlets.
\newblock In \emph{Advances in Neural Information Processing Systems}, pages
  5036--5044, 2017.

\bibitem[Chouldechova(2017)]{chouldechova2017fair}
A.~Chouldechova.
\newblock Fair prediction with disparate impact: A study of bias in recidivism
  prediction instruments.
\newblock \emph{Big data}, 5\penalty0 (2):\penalty0 153--163, 2017.

\bibitem[Chouldechova and Roth(2018)]{chouldechova2018frontiers}
A.~Chouldechova and A.~Roth.
\newblock The frontiers of fairness in machine learning.
\newblock \emph{arXiv preprint arXiv:1810.08810}, 2018.

\bibitem[Cohen et~al.(2015)Cohen, Elder, Musco, Musco, and
  Persu]{cohen2015dimensionality}
M.~B. Cohen, S.~Elder, C.~Musco, C.~Musco, and M.~Persu.
\newblock Dimensionality reduction for {$k$}-means clustering and low rank
  approximation.
\newblock In \emph{Proc. 47th Annu. ACM Sympos. Theory Comput. {\em(STOC)}},
  pages 163--172, 2015.

\bibitem[Dheeru and Karra~Taniskidou(2017)]{Dua:2017}
D.~Dheeru and E.~Karra~Taniskidou.
\newblock {UCI} machine learning repository, 2017.
\newblock URL \url{http://archive.ics.uci.edu/ml}.

\bibitem[Indyk(2001)]{indyk2001algorithmic}
P.~Indyk.
\newblock Algorithmic applications of low-distortion geometric embeddings.
\newblock In \emph{Foundations of Computer Science, 2001. Proceedings. 42nd
  IEEE Symposium on}, pages 10--33, 2001.

\bibitem[Kohavi(1996)]{kohavi1996scaling}
R.~Kohavi.
\newblock Scaling up the accuracy of naive-bayes classifiers: a decision-tree
  hybrid.
\newblock In \emph{KDD}, volume~96, pages 202--207, 1996.

\bibitem[Makarychev et~al.(2018)Makarychev, Makarychev, and
  Razenshteyn]{makarychev2018performance}
K.~Makarychev, Y.~Makarychev, and I.~Razenshteyn.
\newblock Performance of johnson-lindenstrauss transform for {$k$}-means and
  {$k$}-medians clustering.
\newblock \emph{arXiv preprint arXiv:1811.03195}, 2018.

\bibitem[Moro et~al.(2014)Moro, Cortez, and Rita]{moro2014data}
S.~Moro, P.~Cortez, and P.~Rita.
\newblock A data-driven approach to predict the success of bank telemarketing.
\newblock \emph{Decision Support Systems}, 62:\penalty0 22--31, 2014.

\bibitem[R{\"{o}}sner and Schmidt(2018)]{rosner18privacy}
C.~R{\"{o}}sner and M.~Schmidt.
\newblock Privacy preserving clustering with constraints.
\newblock In \emph{45th International Colloquium on Automata, Languages, and
  Programming, {ICALP} 2018, July 9-13, 2018, Prague, Czech Republic}, pages
  96:1--96:14, 2018.

\bibitem[Schmidt et~al.(2018)Schmidt, Schwiegelshohn, and
  Sohler]{schmidt2018fair}
M.~Schmidt, C.~Schwiegelshohn, and C.~Sohler.
\newblock Fair coresets and streaming algorithms for fair {$k$}-means
  clustering.
\newblock \emph{arXiv preprint arXiv:1812.10854}, 2018.

\bibitem[Strack et~al.(2014)Strack, DeShazo, Gennings, Olmo, Ventura, Cios, and
  Clore]{strack2014impact}
B.~Strack, J.~P. DeShazo, C.~Gennings, J.~L. Olmo, S.~Ventura, K.~J. Cios, and
  J.~N. Clore.
\newblock Impact of hba1c measurement on hospital readmission rates: analysis
  of 70,000 clinical database patient records.
\newblock \emph{BioMed research international}, 2014, 2014.

\end{thebibliography}
